\documentclass[aps,pra,preprint,showpacs,groupedaddress,amsmath,amssymb,amsfonts]{revtex4-1}
\usepackage{graphicx}
\usepackage{bm}

\begin{document}


\title{Potential in the Schrodinger equation: estimation from empirical data}


\author{J. L. Subias}
\email[e-mail for correspondence to the author:\\]{jlsubias@unizar.es}
\affiliation{Departamento de Ingenieria de Diseno y Fabricacion, Universidad de Zaragoza,\\C/Maria de Luna 3, 50018-Zaragoza, Spain}


\date{February 24, 2020}

\begin{abstract}
A recent model for the stock market calculates future price distributions of a stock as a wave function of a quantum particle confined in an infinite potential well. In such a model the question arose as to how to estimate the classical potential needed for solving the Schrodinger equation. In the present article the method used in that work for evaluating the potential is described, in the simplest version to implement, and more sophisticated implementations are suggested later.
\end{abstract}


\maketitle

\section{Introduction}
The initial study on the quantum model for stock markets \cite{Subi14} suggested that the energy-matter interaction paradigm could be an excellent assumption for the modelling of investor interactions. On the other hand, the estimation of the classical potential would be possible by applying the radioactive decay paradigm. In what follows, we will describe a numerical method based on both paradigms that in its simplest version already provides good predictive results for future price distributions in a stock market. In addition, known patterns of market technical analysis are presented that could be elegantly explained from a quantum-mechanical point of view.
The presentation will be completed with a short overview of market technical analysis, dedicated to the reader who is not familiar with this terminology.
\section{The model in a first approach}
\begin{figure}[tb]
\centering
 \includegraphics*[scale=2.0]{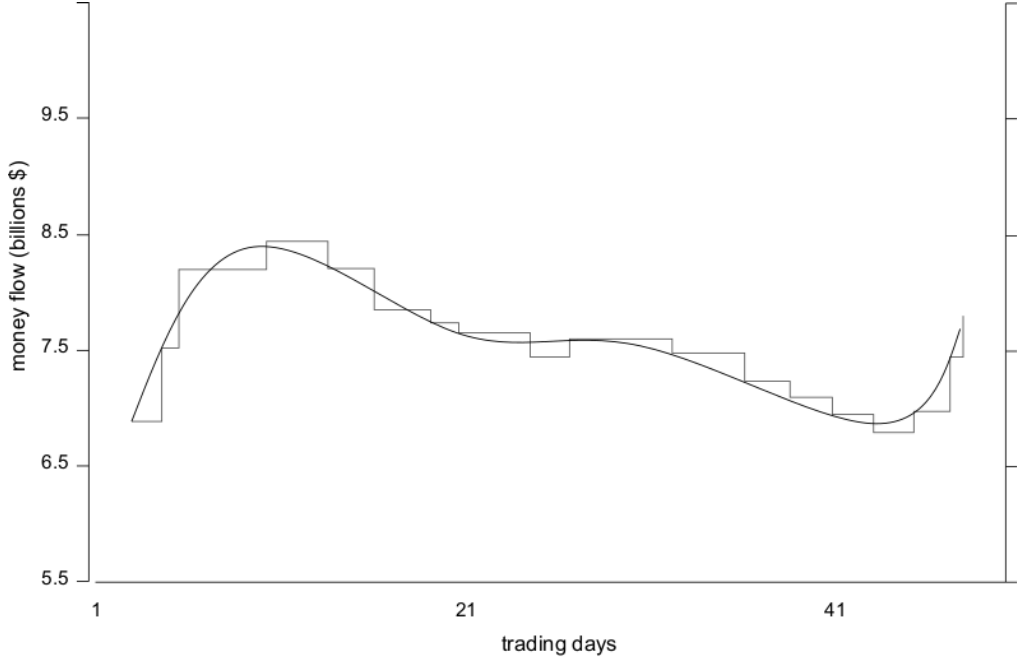}
  \caption{\label{MoneyFlow0}The flow of money appears as a pseudo-trajectory. Its quantum nature is not obvious}
\end{figure}

An investor is similar to an atom that can absorb (buy) or emit (sell) a quantum of energy (block of shares of a certain size). In the "ground" state a investor does not own any stock. If he buys, such an investor goes into an "excited" state. From this state,  investor can sell, returning to a less excited state (reducing or underweighting investment) or to  ground state (total sale of investment).
If a market were made up of a few investors, variations in energy (money flow) would be discrete (see Fig.~\ref{MoneyFlow0}) and quantum nature of transactions would be obvious. In practice, a large number of investors and some kind of decoherence determine that prices take the form of pseudo-trajectories, although totally discontinuous and erratically evolving.


\begin{figure}[tb]
\centering
 \includegraphics*[scale=.7]{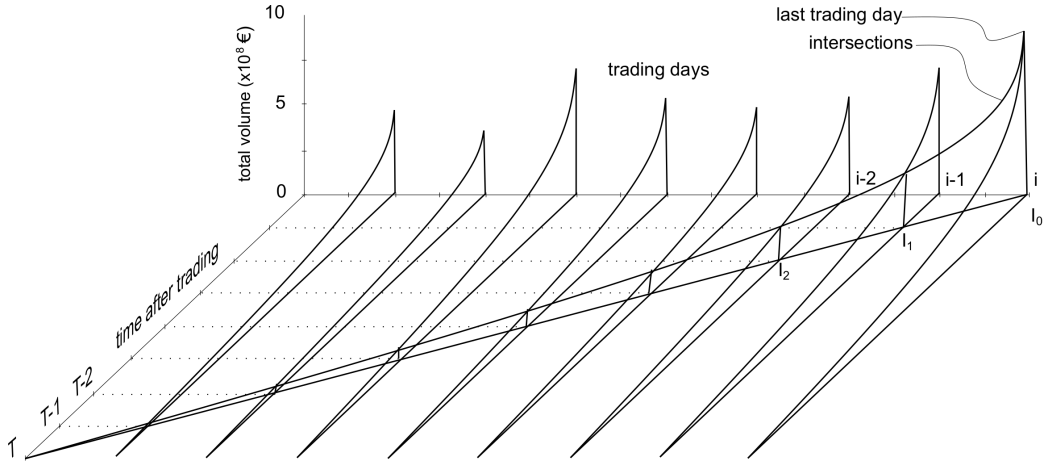}
  \caption{\label{DRadiactiva}Radioactive decay paradigm: global distribution is an intersection}
\end{figure}

For modeling this process, we propose to take as a paradigm the radioactive decay. Thus, the mean life span of an excited state will be globally characterized by a disintegration constant. Thus, whether $N_{0i}$ is the volume traded on $i-th$ day, this means that a number $N_{0i}$ of shares have changed hands on that day. Over time those shares gradually change hands again, so that after a time of $t$ days, the number of remaining shares, not traded (held), will be $N(t)$, approximately fulfilling the known relationship:

\begin{equation}\label{Ni}
    N_{i}(t)=N_{0i}\textrm{e}^{-\lambda t}
\end{equation}
the disintegration constant mentioned above being $\lambda$. Over time, daily trading volumes would evolve as in Fig.~\ref{DRadiactiva}

\begin{figure}[tb]
\centering
 \includegraphics*[scale=1]{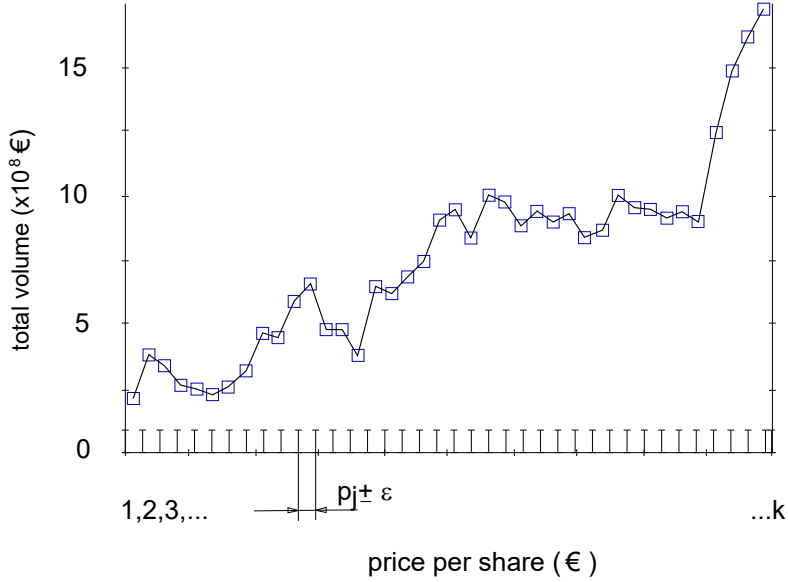}
  \caption{\label{DivCampo}Global distribution of volumes with respect to prices}
\end{figure}

Intersections $I$ are the number of non-traded (held) shares distributed by price and trading days. The probability of a share changing hands can be empirically estimated as the probability of trading shares belonging to the \emph{free float}, i.e:
\begin{equation}\label{Prob}
    P=\frac{v_{m}}{v_{ff}}
\end{equation}
with $v_{m}$, $v_{ff}$ being mean daily volume and free float of the company.
The disintegration constant can be deduced from ~(\ref{Prob}):
\begin{equation}\label{Lambda}
    \lambda(P)=-\ln(1-P)
\end{equation}
the width of potential well will constitute the variation domain of prices $p$, that is, $p_{min}\leq p\leq p_{max}$ . Let us consider this domain divided into $k$ intervals with a width of $2\varepsilon$, centered on price values $p_{j}$, $j=1...k$ (see Fig.~\ref{DivCampo}).

\begin{figure}[tb]
\centering
 \includegraphics*[scale=.7]{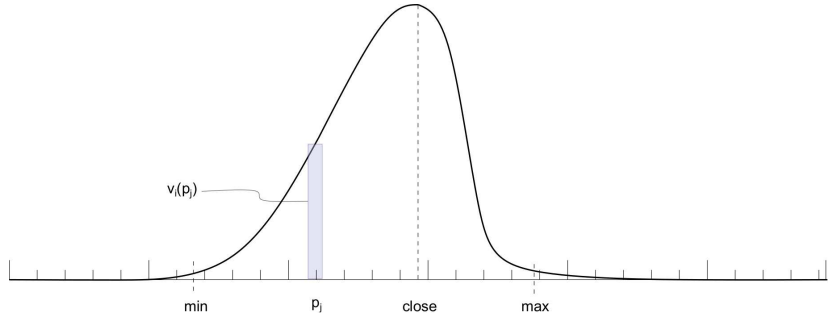}
  \caption{\label{partoGauss}Intraday distribution of volumes $v_{i}$ with respect to traded prices $p_{j}$. The total volume associated with $p_{j}$ is a summation as successive trading days overlap in price}
\end{figure}

In addition, traded volume $v_{i}$ corresponding to $i-th$ day is distributed between a maximum and minimum prices of that $i-th$ day by a proper distribution, being adjusted as shown in Fig.~\ref{partoGauss}.

This step can be simplified by considering the total volume $v_{i}$ associated with the closing price $p_{iclose}$ in $i- th$ trading day, but the probabilistic distribution is a better approximation. The numbers of (held) shares remaining after each trading day would be: $v_{0}\textrm{e}^{-\lambda T}$, $v_{1}\textrm{e}^{-\lambda(T-1)}$,...,$v_{i-2}\textrm{e}^{-2\lambda}$, $v_{i-1}\textrm{e}^{-\lambda}$, $v_{i}$   and the total number corresponding to price $p_{j}$

\begin{equation}\label{Npj}
    N(p_{j})=\sum^{T}_{i=0}v_{i}(p_{j})\textrm{e}^{-\lambda (T-i)}
\end{equation}
$v_{i}(p_{j})$ being the fraction of $v_{i}$ traded at price $p_{j}$.
When quantum particle passes through a certain $p_{j}$ position, the probability of interaction is proportional to the total number of remaining shares for price level $p_{j}$, so we will assume that the potential field would be made up of $V(p_{j})$ values, such that:

\begin{equation}\label{Vpj}
    V(p_{j})\backsim\sum^{+\varepsilon}_{-\varepsilon}p_{\varepsilon}N(p_{\epsilon}),
\end{equation}
being $p_{\varepsilon}\in(p_{j}-\varepsilon,p_{j}+\varepsilon)$.
As we do not know initial configuration $\{V_{0}(p_{j})\}$, we will assume that a flat configuration tends to real configuration for a long enough time. We will set this time frame $T$ equal to the \emph{free float} rotation period, that is, an elapsed time during which a total volume equal to the company's \emph{free float} is traded.
\section{Some examples}
Under previous hypotheses we can obtain several examples that illustrate basic concepts used by technical analysts, namely support levels, resistance, sideway movement,...explained from a quantum-mechanical point of view.
\subsection{Sideway trend}
\begin{figure}[tb]
\centering
 \includegraphics*[scale=1.5]{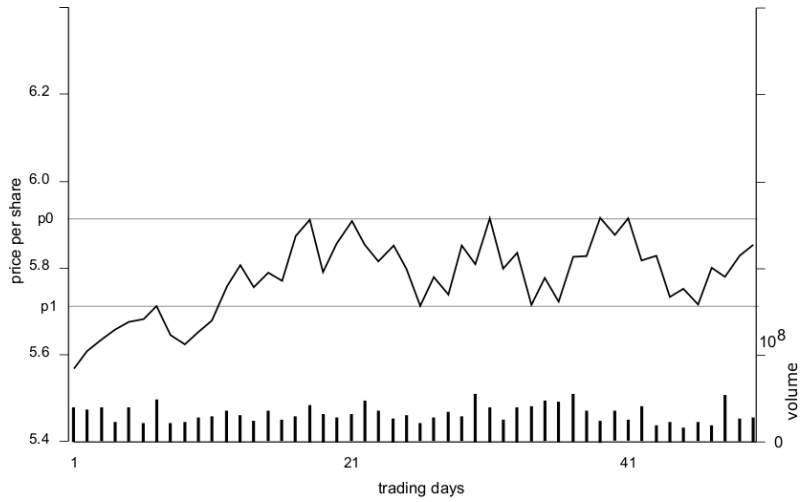}
  \caption{\label{CanalH}price levels $p_{0}$, $p_{1}$ corresponding to the potential peaks}
\end{figure}

In particular, Fig.~\ref{TendLat} is a configuration of potential in which two positions that act as support and resistance (see Appendix) can be clearly seen. Prices tend to move between $P_{0}$, $P_{1}$ in what analysts call a sideway trend, illustrated in Fig.~\ref{CanalH}

\begin{figure}[tb]
\centering
 \includegraphics*[scale=.7]{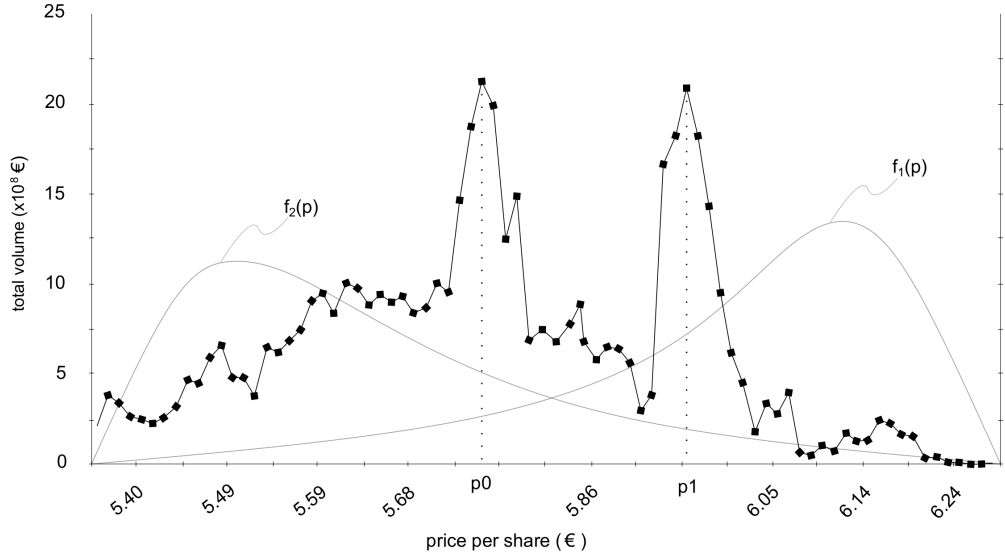}
  \caption{\label{TendLat}Two potential peaks determine sideway trend}
\end{figure}

In sideway trend an analyst's interest is focused on predicting direction (up or down) in which prices will break the $p_{1}-p_{0}$ channel. From our quantum-mechanical point of view, the highest probability of a breakout will be marked by a \emph{wave function}. Thus, if wave function is $f_{1}(p)$, the higher probability corresponds to an upward break, while if it is $f_{2}(p)$, a downward break is more likely.
\subsection{Trend lines}
\begin{figure}[tb]
\centering
 \includegraphics*[scale=.7]{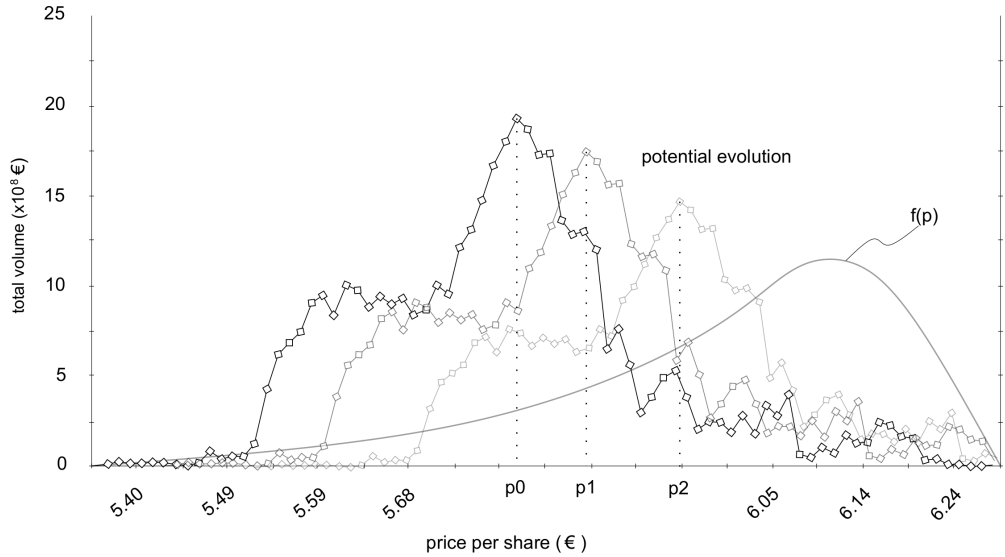}
  \caption{\label{TendAlci}The evolution of potential to determine the upward trend}
\end{figure}

An important concept widely used by analysts is \emph{trend line} (see Appendix) which can be interpreted in quantum-mechanical terms. Thus, in Fig.~\ref{TendAlci}a field of potential with a peak at $p_{0}$ is shown.

\begin{figure}[tb]
\centering
 \includegraphics*[scale=1.5]{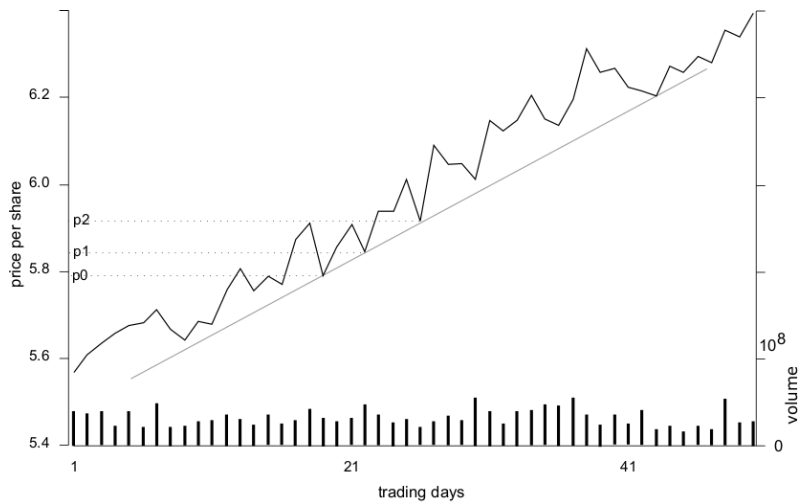}
  \caption{\label{LineAlci}Trend line corresponding to figure \ref{TendAlci}}
\end{figure}

This potential would evolve over time in one way or another, depending on the position occupied by the wave function $f(p)$. For example, if the maximum of wave function is located to the right of $p_{0}$, the potential would migrate to the right, determining the upward trend of Fig.~\ref{LineAlci}.

Correlatively, if $f(p)$ is found to the left of $p_{0}$ we would have the downward trend of Fig.~\ref{LineBaji}.

\begin{figure}[tb]
\centering
 \includegraphics*[scale=.7]{figTendBaji}
  \caption{\label{TendBaji}The evolution of potential to determine the downward trend}
\end{figure}
Thus, the maximum $f(p)$ would be "constructor" and the tail would be "destructor", which would determine the migration of the potential.
\begin{figure}[tb]
\centering
 \includegraphics*[scale=1.5]{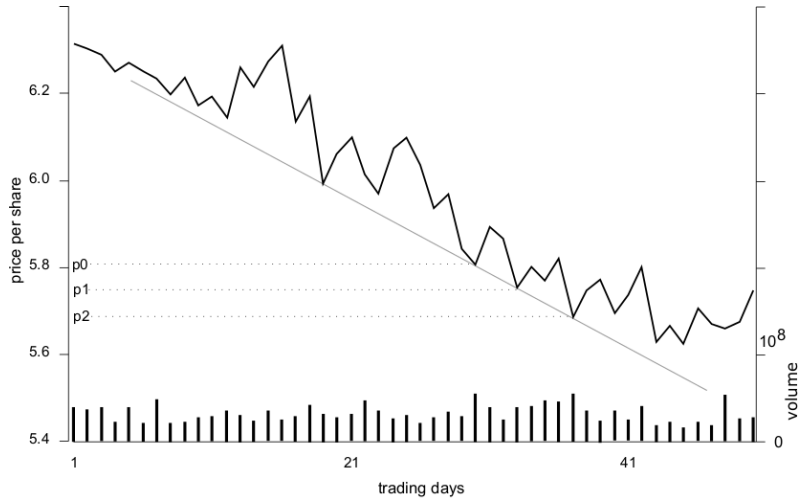}
  \caption{\label{LineBaji}Trend line corresponding to figure \ref{TendBaji}}
\end{figure}
\subsection{Quantum tunneling}
\begin{figure}[tb]
\centering
 \includegraphics*[scale=1.5]{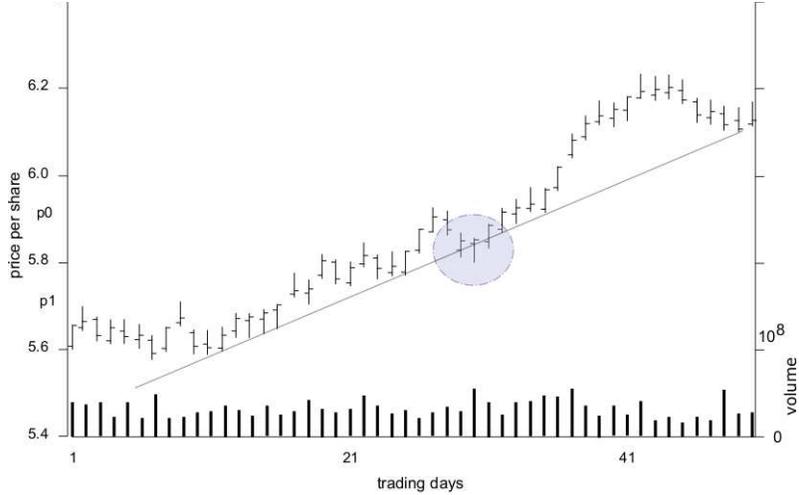}
  \caption{\label{Tunel}Presumed quantum tunneling with a relatively high reflection factor}
\end{figure}

\begin{figure}[tb]
\centering
 \includegraphics*[scale=1.5]{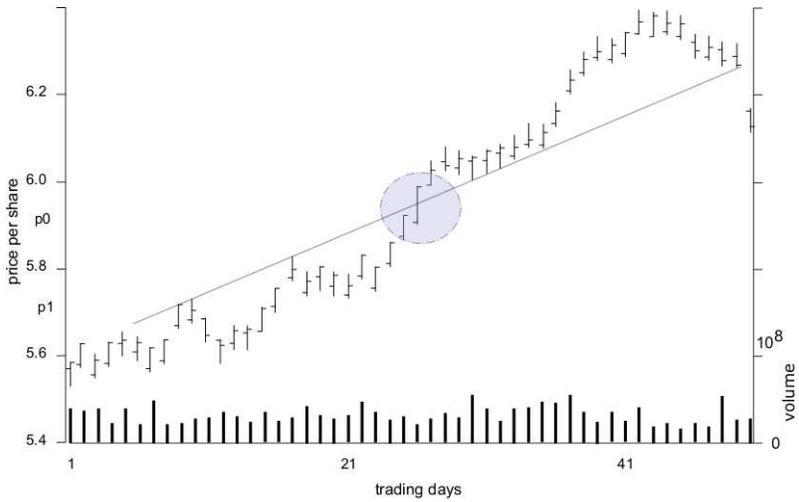}
  \caption{\label{Tunel2}Presumed quantum tunneling with a relatively high transmission factor}
\end{figure}

A figure well known by analysts is the intersection of prices trajectory with a trend line, which is either exceeded upwards or downwards, and which finally either bounces (Fig.~\ref{Tunel})or breaks (Fig.~\ref{Tunel2})the trend.

A possible interpretation of the phenomenon in quantum-mechanical terms would be a hypothetical \emph{quantum tunneling effect} \cite{Raco01} in which the quantities of \emph{transmission} and \emph{reflection} factors would determine a rupture or rebound of quotes.
\section{A more elaborated model}
\begin{figure}[tb]
\centering
 \includegraphics*[scale=.7]{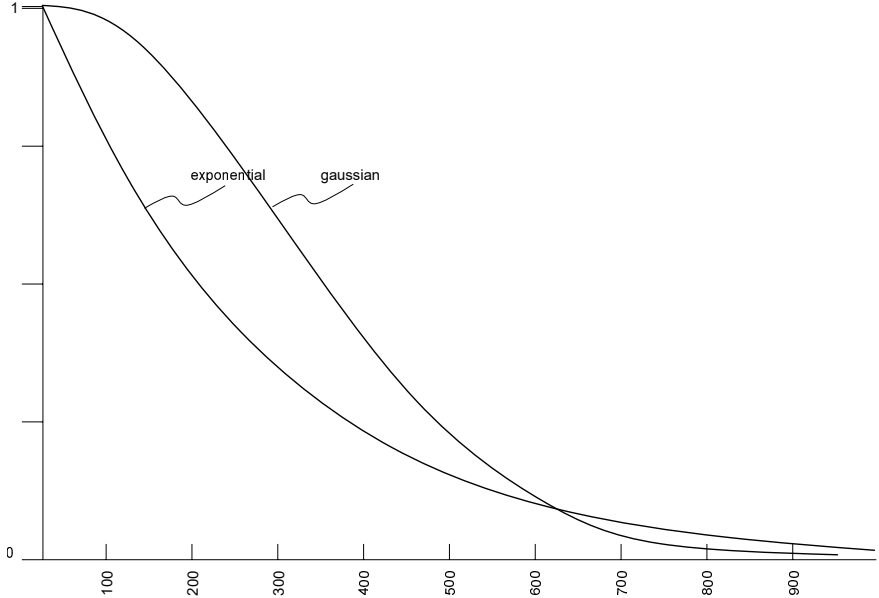}
  \caption{\label{ExpoGauss}Gaussian versus exponential model}
\end{figure}

Daily variations in trading volume clearly indicate that $\lambda$ is not strictly constant, but oscillates around an average value. This average value would be used in the first approximation model described above. On the other hand, it is known that volume is related to volatility, which suggests a dependency such as $\lambda= G(1/\sigma)$, with $G$ probably being a linear function. Therefore, a more elaborate model would be one that contemplates the following dependence:

\begin{equation}\label{NeG}
    N_{i}(t)=N_{0i}\textrm{e}^{-G(1/\sigma) t}
\end{equation}
which would give rise to a family of daily curves dependent on parameter (daily volatility) $\sigma$. This type of exponential model is characterized because the tangent in $t = 0$ has a pronounced slope (see Fig.~\ref{ExpoGauss}),

i.e., the assumption that investors start selling on next day after they have bought. If we wanted the model to reflect a tendency of investors to hold they block of shares for a certain period, we would have to use a Gaussian rather than an exponential model of type~(\ref{fi})

\begin{equation}\label{fi}
    \phi(x)=\frac{1}{\sigma\sqrt{2\pi}}\textrm{e}^{-(t/(\sqrt{2}\sigma))^{2}}
\end{equation}
The tendency to hold blocks of shares is a characteristic that depends on type of market considered and, in particular, percentage of prevalent short-term trading.
\section{Conclusions}
This article describes a method for estimating the potential in the Schrodinger equation for a stock market. By inserting this potential into the infinite well quantum model, such a model provides a future price distribution as a wave function of a quantum particle confined into well.
The first results obtained suggest elegant quantum-mechanical explanations of various patterns belonging to the so-called \emph{technical analysis} of the stock markets.
\appendix*
\section{Technical Analysis Basics}
\subsection{Data and variables}
\begin{figure}[tb]
\centering
 \includegraphics*[scale=1.5]{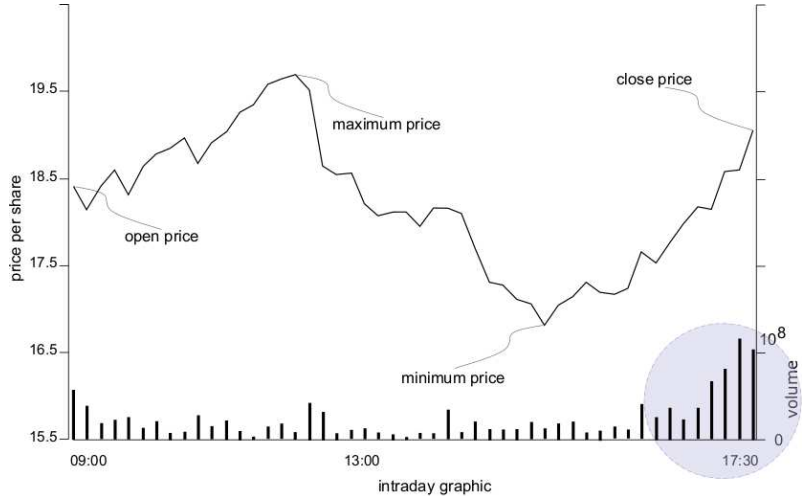}
  \caption{\label{intradia}A typical trading day.  Note that the volumes increase at the end of the day}
\end{figure}

The fundamental variables used in technical analysis are price and volume, the latter being the number of shares traded during a trading day. The market, at the end of the day, offers the data in EOD (End Of Day) format as follows:
 date, opening price, maximum price, minimum price, closing price (see Fig.~\ref{intradia}).

All these data (historical data) are usually offered in the form of a CSV (Comma Separated Values) file on main financial websites, some of which are free of charge.
\subsection{The trend}
\begin{figure}[tb]
\centering
 \includegraphics*[scale=1.5]{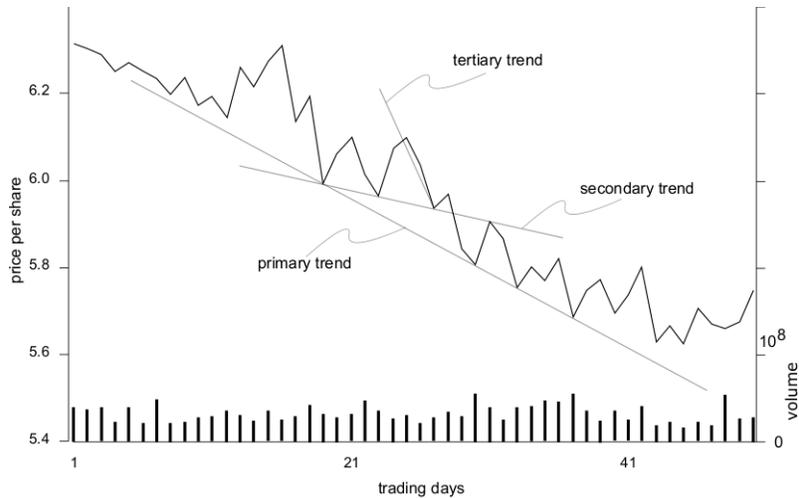}
  \caption{\label{LineBaji2}The three main trend categories: this suggests a fractal structure}
\end{figure}

The trend is the predominant direction that prices take as a result of the evolution of supply and demand. If for a particular asset the balance between supply and demand were perfect, asset prices would be constant over time. Usually this is not the case and three types of trend appear, namely
\begin{itemize}
  \item upward trend, when demand exceeds supply the buyers drive the price up.
  \item otherwise, in a downward trend sellers push the price down.
  \item sideway trend occurs when there is no predominant direction of price movement.
\end{itemize}
For some reason, the evolution of supply and demand is approximately linear over time, so it is possible to graphically represent trends using straight lines (see Fig.~\ref{LineBaji2}).

According to their duration the trends are classified in:
\begin{itemize}
  \item Primary: more than 11 months
  \item Secondary: 2 to 11 months
  \item Tertiary: less than 2 months
\end{itemize}
The empirical observation that duration and slope of a trend are correlated is very interesting, and thus the greater the slope, the weaker the trend and the greater probability of being broken. The most stable trends are those between 30 and 45 degrees of inclination. The relative strength of a trend depends on its duration and number of bounces it has sustained.
The primary trend is the one that defines the state of the market:
\begin{itemize}
  \item Bull market, bullish (upward primary trend)
  \item Bear market, bearish (downward primary trend)
\end{itemize}
\subsection{Supports and resistances}
\begin{figure}[tb]
\centering
 \includegraphics*[scale=1.5]{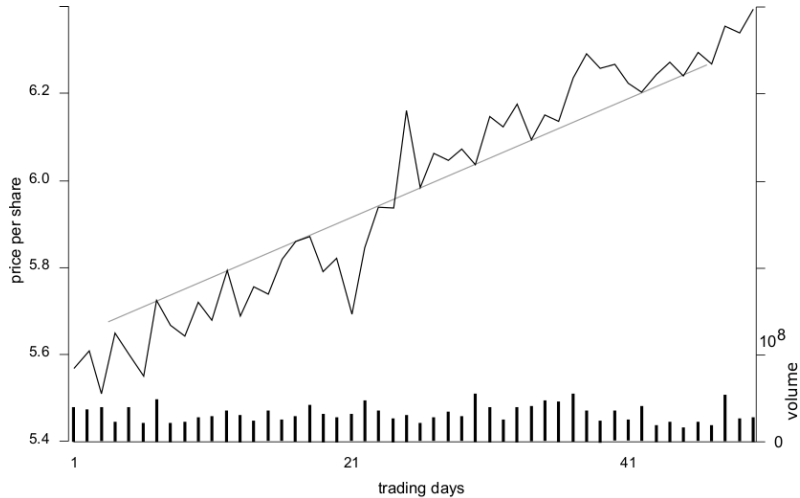}
  \caption{\label{resisopor}When a resistance is overcome, it becomes a support}
\end{figure}

\begin{figure}[tb]
\centering
 \includegraphics*[scale=1.5]{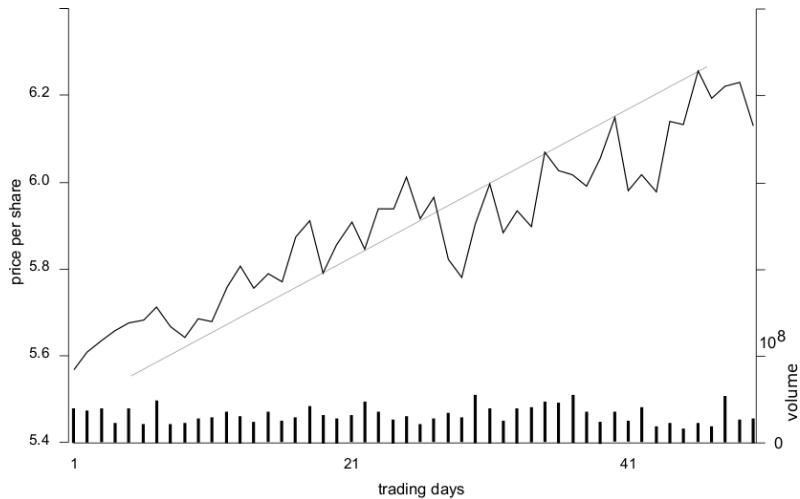}
  \caption{\label{soporresi}When a support is broken, it becomes a resistance}
\end{figure}

A trend line that passes through price trajectory highs is called support. Correlatively, a line resulting from joining lows is called resistance. Supports and resistances tend to act as barriers, preventing prices from move to a certain direction.
Support or resistance level is also said to refer to a particular price that is difficult to drill down or break up.
When a resistance is broken upwards, it becomes support (Fig.~\ref{resisopor}), and vice versa, a support in resistance (Fig.~\ref{soporresi})

\begin{figure}[tb]
\centering
 \includegraphics*[scale=1.5]{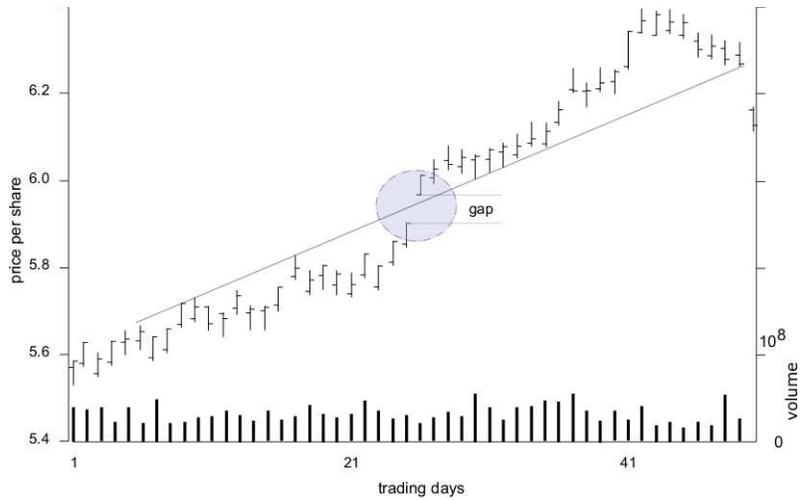}
  \caption{\label{gap}A gap: interesting phenomenon to explain in quantum-mechanical terms}
\end{figure}

 A very significant type of drilling or breakage is when it takes place with an upward or downward \emph{gap}. A gap is a price range in which no shares are traded and can only be seen well on bar charts (see Fig.~\ref{gap})

\begin{figure}[tb]
\centering
 \includegraphics*[scale=1.5]{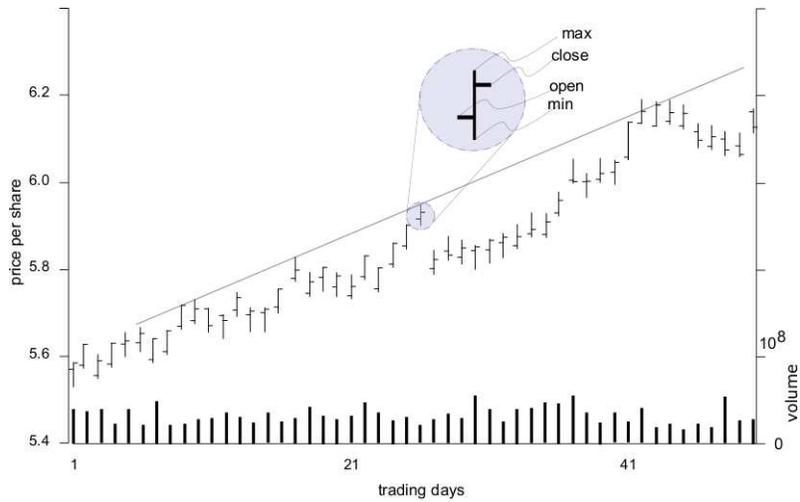}
  \caption{\label{grafBarras}Bar chart showing intraday data}
\end{figure}

The price path can be represented by a single line formed by the closing prices. But usually it is represented by bars that reflect the EOD (End Of Day) values opening, maximum, minimum and closing (see Fig.~\ref{grafBarras})

\subsection{Channels}
\begin{figure}[tb]
\centering
 \includegraphics*[scale=1.5]{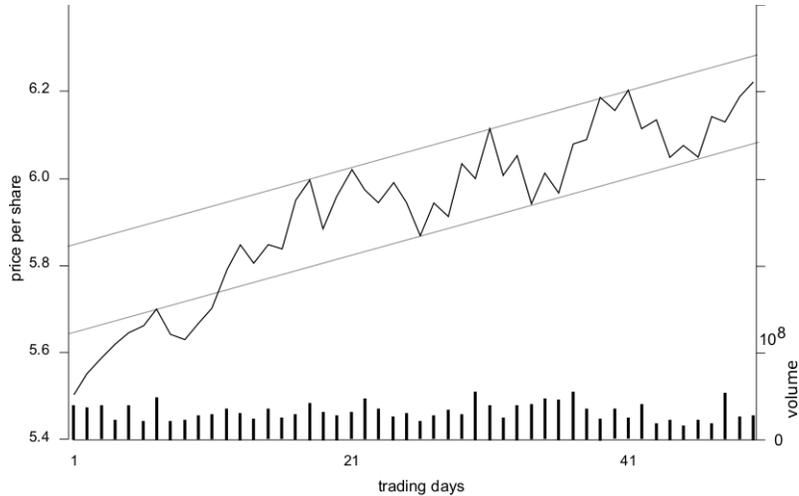}
  \caption{\label{Canal}Channel: a trend being characterized by parallel support and resistance}
\end{figure}

They occur when the price path moves between a parallel support and resistance (see Fig.~\ref{Canal}).
In a sideway trend, horizontal channels are often formed (see Fig.~\ref{CanalH}).
\subsection{Triangles and wedges}
\begin{figure}[tb]
\centering
 \includegraphics*[scale=1.5]{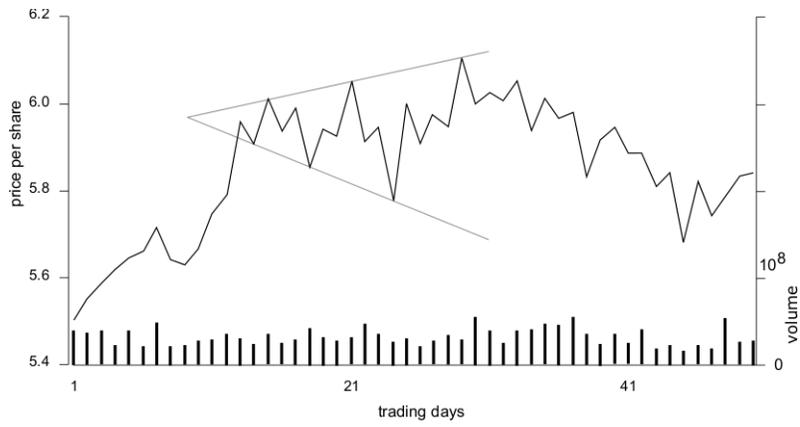}
  \caption{\label{TrianExp}Expanding triangle}
\end{figure}

\begin{figure}[tb]
\centering
 \includegraphics*[scale=1.5]{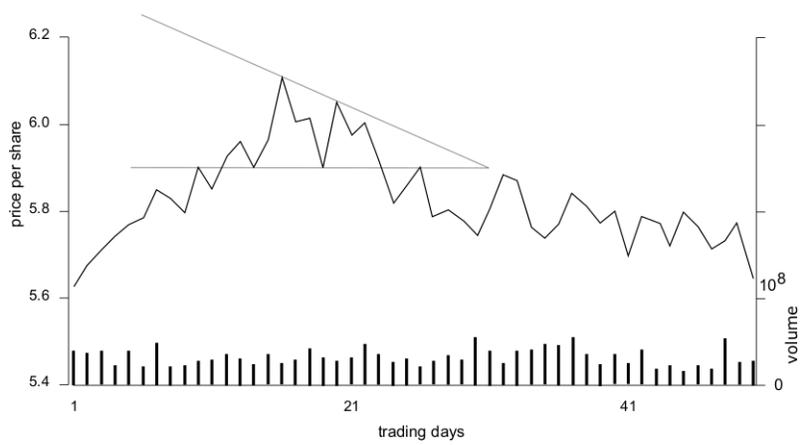}
  \caption{\label{TrianH}The outcome of a triangle depends on the relative strength of its trend lines}
\end{figure}

They occur when support and resistance lines intersect. Whether the slope of the two lines has the same sign or not determines its designation. Usually triangles and wedges are convergent, but sometimes they are expanding or divergent (see Fig.~\ref{TrianExp}).
Since the relative strength of the trend lines correlate with their slope, it is sometimes possible to predict how the break will be. Example: the horizontal line is always weaker than a line that has a 30-45 slope degrees, therefore, in see Fig.~\ref{TrianH} one can expect the break to occur downwards.

\begin{figure}[tb]
\centering
 \includegraphics*[scale=1.5]{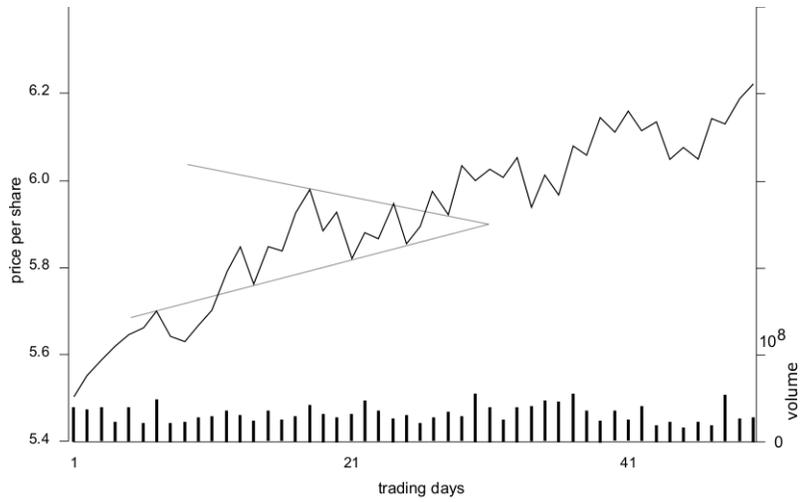}
  \caption{\label{Triangulo}Triangle: lines of support and resistance with opposite slopes}
\end{figure}
\begin{figure}[tb]
\centering
 \includegraphics*[scale=1.5]{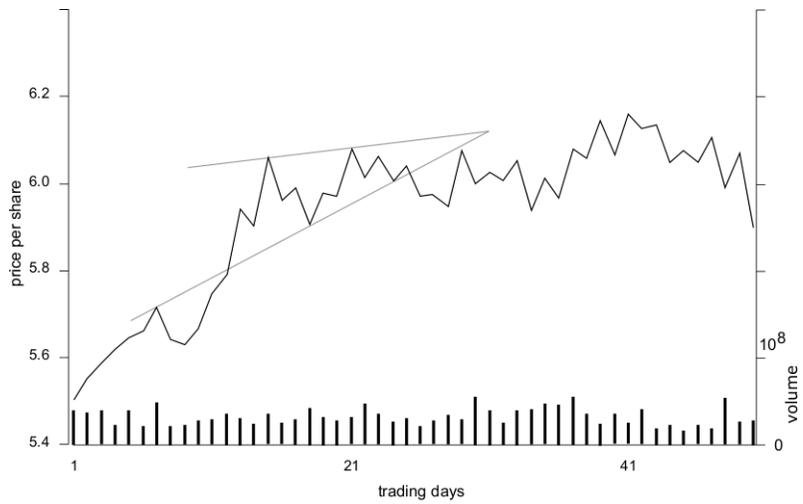}
  \caption{\label{Cunia}Wedge: support and resistance with slopes of the same sign}
\end{figure}

\bibliography{../SpinModel}

\end{document}